\documentclass{emulateapj}
\usepackage{amsmath}

\shorttitle{Evolution of radio galaxies}
\shortauthors{Luo \& Sadler}

\begin{document}

\title{The evolution of extragalactic radio sources}

\author{Qinghuan Luo and Elaine M. Sadler}
\affil{School of Physics, University of Sydney, NSW 2006, Australia}

\begin{abstract}
A model for the evolution of low-luminosity radio galaxies is presented. In the model,
the lobes inflated by low-power jets are assumed to expand in near pressure-balance
against the external medium. Both cases of constant external pressure and decreasing
external pressure are considered. Evolution of an individual source is described by 
the power-size track. The source appears as its lobe is inflated and radio luminosity 
increases to above the detection level; the source then moves along the track and eventually disappears as
its luminosity drops below the detection limit. The power-size tracks are calculated including the 
combined energy losses due to synchrotron radiation, adiabatic expansion,
and inverse Compton scattering. It is shown that in general, the constant-pressure 
model predicts an excess number of luminous, small-size sources while underpredicting 
large-size sources in the power-size diagram. The predicted spectra are steep for most 
sources, which is inconsistent with observations. By comparison, the pressure-limiting model 
fits observations better. In this model, low-luminosity sources undergo substantial expansion losses in 
the initial phase and as a result, it predicts fewer luminous, small-size sources. 
The resultant spectra are flat for most sources except for the oldest ones, which
seems consistent with observations. The power-size tracks, in contrast to
that of high-luminosity radio galaxies, are characterized by a slow increase in luminosity 
for most of the source's life, followed by a rapid decline when the synchrotron 
or inverse Compton scattering losses set in.

\end{abstract}


\section{Introduction}

The generic model for radio galaxies assumes that twin jets emanating 
from an active galactic nucleus propagate outward in two opposite directions.
The jets, which initially propagate at a relativistic speed, interact
with the surrounding medium leading to formation of a diffuse emission region. 
Radio galaxies appear to have two classes: low- and high-luminosity radio
galaxies, commonly referred to as FR I and II sources, respectively \citep{fr74}. 
The jets in high-luminosity radio galaxies have relatively homogeneous 
morphology; they are well collimated and propagate
through the surrounding medium---initially in the cores, then halos of their 
parent galaxies and then the intergalactic medium (IGM)---creating 
pair of large lobes. The jets are dim until the end of the lobes where there are bright hot spots. 
Classical double radio sources are a typical example of this class.

By contrast, low-luminosity radio galaxies are characterized by
jets that are bright close to the nucleus of their parent galaxy. 
The jets have diverse morphologies, a feature that can be interpreted as deceleration 
of jets due to entrainment of the external medium. The jets are initially 
laminar near the nucleus and then subject to turbulent disruption
when passing through the flare region that is thought to be the main acceleration
site for relativistic particles. The jets beyond the flare region spread out,
resembling smoke arising from a chimney mixing with the ambient medium. 

The key issues in the understanding of radio galaxies include 
the evolution of radio galaxies and the underlining physics that distinguishes 
these two classes. One suggestion is that these two classes of source 
are intrinsically different, primarily in their jet dynamics,
evolving along different tracks \citep{jw99}. However, there 
are suggestions that some of the high-luminosity radio sources with weak jets may evolve into 
low-luminosity sources \citep{gw87,gw88,kb07}. 
Some radio sources exhibit mixed features of FR Is and IIs. 
For example, there are souces with one-side jet showing the FR I features and the other showing
the FR II features. This leads to an opinion that such classification may not be 
clear cut as previously thought \citep{kb07}. 

It is well accepted that the radio emission in radio galaxies
is due to synchrotron radiation by relativistic electrons (or positrons) 
injected from the jets.  The total synchrotron power $P_\nu$ evolves with time as the 
injection of a mixture of kinetic energy and magnetic energy competes against
the losses due to volume expansion and radiation. When the losses dominate, the total power 
is a decreasing function of time as the source ages. Since the typical evolutionary
time scale is $\sim 10^8\,\rm yr$, it is not practical to measure how the total power changes
in time directly by observations. One may study the temporal evolution of radio galaxies 
from the total spectral power, $P_\nu$, as a function of the source's linear size \citep{s63}. 
The linear size here is defined as the dimension
of the lobe along the jet axis. Since the linear size $D$ increases as the 
source expands, a radio source should evolve along a particular track
in the $P_\nu$--$D$ diagram. 

There are many discussions in the literature on 
the time evolution of high-luminosity radio galaxies (or FR II sources)
\citep{ketal97,betal99,mk02}. 
In the existing models, there are three relevant regions where
the physical processes determine the evolution of the source. These include the hot spots 
where particles are assumed to be accelerated and radiate, the head region that
contains the hot spots, and the lobe---an emission volume inflated by the input of the
jet, where relativistic particles are injected and the volume of the emitting plasma expands.
Since the hot spots appear at the end of the jet, their distance to the center can be identified as the 
linear size of the lobe. Thus, the modeling of the time evolution of the lobe size 
is reduced to the problem of modeling of changes in the location of the hot spots. One of the 
widely-discussed models is the self-similar expansion model in which the jet 
creates a bow shock by sweeping up the ambient material  \citep{fz86,f91,kf98}. 
So, the location of the shock is completely determined by the jet power and the density of 
the surrounding medium. This allows one to establish the source size as a function of time.  

In this paper we consider the evolution of low-luminosity radio galaxies
(or FR Is). We derive the radio power as a function of the source's size, $P_\nu(D)$,
which can be directly compared with the power-size diagram inferred from 
observations. In contrast to FRIIs, there are few discussions on the evolution of the
low-luminosity sources, mainly because of the lack of quantitative models 
that relate the linear sizes to the jet dynamics. 
One of the defining features of FR Is is that the flare region, the hot-spot 
equivalent as compared to high-luminosity sources, is located close to the nucleus, which
indicates that the jet is decelerated to the subsonic flow regime 
relatively close by the nucleus. Since the jet continues to expand well beyond the flare
region, forming a diffuse emission region further beyond, the size of a FR I source 
is not directly related to the location of the flare region and should be determined separately from
the expansion of the diffuse region. Here we continue to refer to this diffuse 
emission region as the lobe as in FR IIs despite the significant difference between 
their morphologies. For low-power jets, the physical conditions of
the surrounding medium play a critical role in defining the change in the lobe size. 
If the medium is warm, the expansion proceeds at near pressure balance against the 
ambient pressure and the volume increases more slowly with time than the self-similar 
expansion in the high-luminosity sources.  As a result, the low-luminosity sources
grow in size, by comparison, much more slowly than the high-luminosity sources.

In Sec 2, we discuss a generic model for both high-power and low-power
jets. The evolution of the emitting plasma in the lobes is considered in
Sec 3, with application to high-power jets in Sec 4. The evolution of 
low-luminosity radio sources is discussed in Sec 5. 

\section{Generic model for jet-fed radio sources}

\begin{figure}
\includegraphics[width=8cm]{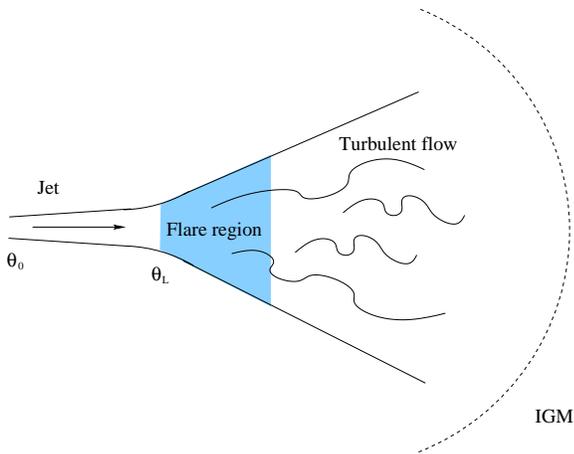}
\caption{A schematic diagram for a radio source fed by a low-power jet. The flare region has 
an opening angle $\theta_L$ much larger than the jet opening angle $\theta_0$. The dashed
line indicates the outer boundary of the diffuse emission region that separates it from 
the external medium, i.e. the galaxy halo or intergalactic medium (IGM). In contrast to
high-power jets, no shock develops at the boundary.
}
\label{fig:frjet}
\end{figure}

We consider a generic model of extragalactic radio sources, in which
a relativistic jet emanates from the central engine, inflating a large diffuse 
radio emission region. In the current models of FR IIs \citep{betal99}, 
it is generally assumed that particles are accelerated by shocks. The 
accelerated particles diffuse through the surrounding region, referred 
to as the head region where hot spots are located, and are injected into
the lobe. One assumes that these basic ingredients are also true for FR Is (cf. Sec 5)
except that for low-power jets the shocks occur close to the nucleus and the jets beyond the shocks 
continue to decelerate.  A schematic diagram for a low-power jet is shown in figure \ref{fig:frjet}. 
The region where shocks occur is referred to as the flare region.
In observations, the bright spots in the flare region can be regarded as 
the hot spot counterparts when compared to high-luminosity sources.
The diffuse emission region, which extends well beyond the shock, is regarded
as the lobe in analogy with FR IIs.

\subsection{Self-similar expansion of jets}

For radio sources driven by high-powered jets, the effect 
of the pressure of the surrounding medium
on the expansion of the lobes is negligible. As a result, the jet-driven lobe expands in 
a self-similar manner. The distance from the central engine to the shock 
can be derived using the self-similar expansion argument \citep{fz86,f91,kf98}. 
It is convenient to define a working surface at the shock, separating the 
jet from the lobe. Such hypothesis may appear to be oversimplified, especially for
multiple shocks that may occur, but it is useful in formulating a global model
that links the various components of the jet-lobe system based 
energy conservation.  The input power by the jet across the surface is 
$Q_j$ and remains constant over the lifetime, where $Q_j$ is the jet power. 
The work done by the jet against the ram pressure is $W\sim \rho\theta^2v^2r^3_s$, 
where $r_s$ is the radial
distance to the surface, $v\sim \dot{r}_s$ is the advancing speed of the jet and
$\theta$ is the opening angle of the jet. The density profile of the ambient medium,
i.e. the halo of its parent galaxy, can be modeled as $\rho=\rho_c(1+r^2/r^2_c)^{-\beta/2}$, where 
$\rho_c\sim 1.7\times10^{-23}{\rm kg}\,{\rm m}^{-3}$ is the core density and $r_c$ is
the core radius, typically about a few kpc \citep{gc91,mz98,betal99}. The profile index 
$\beta$ is between 1 and 2.5. Writing $W\sim Q_jt$, one obtains $r_s$ 
as a function of time:
\begin{equation}
r_s(t)\approx c_1\left({Q_j\over\rho_cr^\beta_c}\right)^{1/(5-\beta)}t^{3/(5-\beta)},
\label{eq:Dt}
\end{equation}
with $c_1\equiv[(5-\beta)/2\theta]^{2/(5-\beta)}$ and $r>r_c$.
For $r\ll r_c$, one has $r_s(t)\propto t^{3/5}$.

For high-power jets, applicable for FR II sources, $2r_s$ can be identified as the 
linear size of the lobe, denoted here by $D(t)=2r_s(t)$ with $r_s(t)$ given by (\ref{eq:Dt}).
One obtains
\begin{equation}
D(t)=D_0\left({t\over t_0}\right)^{3/(5-\beta)},
\label{eq:Dt2}
\end{equation}
where $D_0=2c_1(Q_j/\rho_cr^\beta_c)^{1/(5-\beta)}t^{3/(5-\beta)}_0$ and $t_0$ is the 
initial time when the size is $D_0$. The initial size $D_0$ is usually set to 
the radius of the hot-spot region. 

\subsection{Pressure-limiting expansion}

The self-similar model considered in Sec 2.1  may not be applicable for 
low-power jets. The pressure in the lobes inflated by the low-power
jets can drop rapidly to about the pressure of the external medium.
Since the density decreases from the core $r>r_c$, the pressure of the 
external medium decreases. We assume that the pressure in the lobes 
continues to balance the external pressure that decreases outward. 
We refer this scenario to as the pressure-limiting expansion. 
The argument for the pressure-limiting expansion can be understood as follows.
Assume that the lobe expansion starts with near pressure balance $p_l\sim p_{\rm ex}$.
If a significant imbalance $p_l>p_{\rm ex}$ develops as a result of substantial drop in 
the external pressure as the lobe expands into the underdense region, 
the expansion will accelerate and the system will relax to near pressure balance 
on a time much shorter than the age of the source.
One can derive the relevant expansion law from the energy conservation
\citep{b86,es89}. Denoting the total 
energy of the lobe by $E_l$, the energy equation is given by
\begin{equation}
{dE_l\over dt}=Q_j-p_l{dV\over dt},
\label{eq:El}
\end{equation}
where one ignores the radiative losses and assumes that the injection power is $Q_j$. 
It should be noted that $p_l$ is sum of all components (radiating particles, nonradiating particles,
magnetic field). Eq (\ref{eq:El}) implies that the main cause for decrease in the internal energy 
in the lobe is the volume work done against the external pressure. 
Assuming that the temperature of the external medium is constant, the external 
pressure can be written as $p_{\rm ex}\sim p_c(r_c/r)^{\beta}$. 
One may write the external pressure at the core as $p_c=n_0k_BT_0\approx 1.4\times10^{-11}\,{\rm Pa}$
for $n_0=\rho_c/m_p=10^5\,{\rm m}^{-3}$ and $T_0=10^7\,\rm K$. 
As we are interested in the pressure-limiting 
expansion with $p_l\sim p_{\rm ex}$, one can substitute $p_l\sim p_c(r_c/r)^{\beta}$
for (\ref{eq:El}) to obtain 
\begin{equation}
D=D_0\left({t\over t_0}\right)^{1/(3-\beta)},
\label{eq:Dt3}
\end{equation}
with 
\begin{equation}
D_0=2r_c\Biggl[\left({\Gamma-1\over r^3_cp_c\chi}\right)\left(
{3-\beta\over3\Gamma-\beta}\right)
Q_jt_0\Biggr]^{1/(3-\beta)}.
\label{eq:Dt3b}
\end{equation}
In deriving (\ref{eq:Dt3}), we assume $E_l=p_lV/(\Gamma-1)$ and $V=\chi r^3$, where 
$\Gamma$ is the adiabatic index of the plasma in the lobe.  The geometry of the lobe 
is described by the parameter $\chi$.  Since there is no strong observational evidence for 
the dependence of the aspect ratio on the size, it is justified to treat $\chi$ as an 
independent parameter. As $\chi=4\pi/3$ corresponds to a sphere, one generally has $\chi\ll4\pi/3$. 
 
\subsection{Constant-pressure expansion}

A special case is the constant-pressure expansion \citep{es89,kb07}.
With $p_{ex}=$const, one obtains 
\begin{equation}
D(t)=D_0\left({t\over t_0}\right)^{1/3},\quad
D_0=r_c\left({Q_jt_0\over3 r^3_cp_{ex}\chi}{\Gamma-1\over\Gamma}
\right)^{1/3}.
\label{eq:Dt4}
\end{equation}
Eq (\ref{eq:Dt4}) implies that the size of the radio sources driven by low-power jets 
grows much more slowly than the sources driven by high-power jets.

\subsection{Radio powers of lobes}

The total radio power as a function of time can be written as 
an integration of the single-particle power, $P_s$, 
over the spatially-integrated, time-dependent particle distribution, 
$N(\gamma,t)$, where $\gamma$ is the particle's Lorentz factor. In practice, 
one may assumes that each particle
emits at the characteristic frequency $\nu_c=(3/4)\nu_B\gamma^2$, 
producing a power spectrum $P_s\delta(\nu-\nu_c)$, where $\nu_B$ is the gyrofrequency
and $P_s$ is the single-particle synchrotron power averaged on the particle's pitch
angle, given by
\begin{equation}
P_s\approx\textstyle{4\over3}\sigma_TcU_B\gamma^2, 
\end{equation}
where $\sigma_T$ is the Thomson cross section and $U_B$ is the magnetic 
energy density. If the pitch angle distribution is maintained in the
isotropic state throughout the evolution, the total spectral power can 
be expressed as
\begin{eqnarray}
P_\nu(t)=\int P_s\delta\left(\nu-\nu_c\right)
N(\gamma,t)d\gamma \approx{P_s\,N(\gamma_*,t)\over2
(\nu_B\nu)^{1/2}},
\label{eq:Pnu}
\end{eqnarray}
where $\gamma_*\equiv(4\nu/3\nu_B)^{1/2}$ is the Lorentz factor of 
particles that emit synchrotron radiation predominantly at frequency 
$\nu$. Eq (\ref{eq:Dt2}) can be used to eliminate
$t$ in (\ref{eq:Pnu}), giving rise to $P_\nu$ as a function of $D$. All the variables in
(\ref{eq:Pnu}) are global, in the sense that they can be regarded as spatial averages. 

One can show that the synchrotron output (in radio) only comprises a tiny fraction of 
the total power input by the jet. The total number of relativistic particles
that emit synchrotron radiation at frequency $\nu$ is $\sim \eta Q_j t\gamma^{-p}_*$,
where $\eta\leq1$ is the fraction of the jet power into the relativistic particles,
$p$ is the particle spectral index and $t$ is the age of the source. Since the 
total synchrotron power is $\sim P_\nu\nu$, the ratio of the synchrotron power to the jet input ($\eta Q_j$)
is estimated as $P_\nu\nu/\eta Q_j\sim \sigma_TcU_Bt(\nu/\nu_B)^{1/2}$. For
$B=50\,\rm nT$, $t=1\,\rm Myr$, and $\nu=1\,\rm GHz$, this ratio is $\sim 10^{-4}$, which 
implies that only a tiny fraction of the jet power is converted to synchrotron radiation.

\subsection{Limiting flux}

The instrument sensitivity places a lower-limit on the observable flux density.
As a result, the $P_\nu$--$D$ tracks have a cut-off at which the flux density is
too low to be observable. To estimate the cut-off, one may write the flux density as
\begin{equation}
F_\nu\sim P_\nu/S\geq F_{\nu *},
\end{equation}
with $F_{\nu *}$ the limiting flux density and $S$ the effective surface area of
the lobe. Hence, the following expression
\begin{equation}
P_\nu\sim F_{\nu *}S,
\end{equation}
defines a cut-off line. The sources below the cut-off line are undetectable
due to the limit of the sensitivity. The simplest case is $S\sim D^2$, which yields
$P_\nu\sim F_{\nu *}D^2$.

\section{Evolution of the emitting plasmas}

The temporal evolution of the relativistic particle spectrum due to
both adiabatic and radiative losses can be derived
by the usual method, i.e. solving a diffusion-loss equation \citep{l94}. 
Since there is no strong evidence for frequency dependence of
the source size, one may ignore spatial diffusion in the treatment of the
evolution of the particle spectrum and only consider the spatially integrated 
distribution $N(\gamma,t)$. Here we
outline the main results based on the solutions discussed in \citet{k62}
and concentrate on the case of the synchrotron losses in magnetic fields that 
decay with time.

\subsection{Time-dependent particle spectra}

When spatial diffusion is neglected, the diffusion-loss equation is simplified to
the usual continuity equation,
\begin{equation}
{\partial N(\gamma,t)\over \partial
t}-{\partial\over\partial\gamma}\left[bN(\gamma,t)\right]=q_l(\gamma,t),
\label{eq:CEq}
\end{equation}
where $q_l(\gamma,t)$ is the particle injection rate,
$b(\gamma)\equiv-\dot{\gamma}$ is the energy loss rate, given by
\begin{equation}
b(\gamma)=\left({\alpha_{_V}\over t}+{\gamma\over\tau_1}
\right)\gamma,\quad\quad \tau_1\equiv{\tau_s\tau_{_{ICS}}\over\tau_s+\tau_{_{ICS}}},
\label{eq:b}
\end{equation}
where $\tau_s=3m_ec^2/(4\sigma_TcU_B)$,  $\tau_{_{ICS}}=3m_ec^2/(4\sigma_TcU_{ph})$, 
$U_{\rm ph}$ is energy density of the seed photons. As usual, one assumes that the injection
rate is time-independent, with a power-law distribution in energy:
\begin{equation}
q_l=q_0\gamma^{-p},\quad \gamma_1\leq\gamma\leq\gamma_m.
\label{eq:ql}
\end{equation}
The injection rate is set to zero outside this range (i.e. $q_l=0$ for
$\gamma<\gamma_1$ and $\gamma>\gamma_m$). For diffusive shock acceleration,
the power index is close to $p\sim 2$.
Assuming that the injection power is $\eta Q_j$ with $\eta<1$ the efficiency for
the jet power going into the relativistic particles, one obtains
$q_0=\eta Q_j/(\langle\gamma\rangle m_ec^2)=
\eta Q_j(p-2)/[m_ec^2(\gamma^{2-p}_1-\gamma^{2-p}_m)]$, where $\langle\gamma\rangle$ is the
average Lorentz factor of the relativistic particles.  
Two simplifications in regarding the continuity equation are made here.
First, the effects of particle pitch angles can be eliminated by assuming 
that the pitch-angle distribution is isotropic and that the isotropic distribution
can be maintained by efficient pitch angle scattering by plasma turbulence in
the lobe. This assumption is valid if the isotropisation time is much shorter 
than the synchrotron cooling time \citep{jp73}. Second, particles with $\gamma<\gamma_1$ are 
ignored in the calculation of the power spectrum.
Due to energy losses, particles can migrate from above the cut-off to below
the cut-off ($<\gamma_1$). Thus, the particle spectrum that is initially zero
below the lower cut-off can become nonzero. We assume that the lower cut-off
is the order magnitude of the bulk Lorentz factor of the jet, typicaly about 
ten. The typical Lorentz factor for high-frequency 
emission is $\gamma_*\sim 10^3(\nu/1\,{\rm GHz})^{1/2}
(50\,{\rm nT}/B_0)^{1/2}$. 
Since low-energy particles do not contribute to the high-frequency emission, we can safely 
ignore the particles with $\gamma<\gamma_1$ in our calculation.
In practical calculations (cf. Sec. 4 and 5), the results are not sensitive to the 
choice of $\gamma_m$ as long as it is well above that required for the maximum 
observing frequency. 

Assuming that a particle is injected at $t_0$ with the initial Lorentz factor
$\gamma_0$, the solution to 
the equation of the single particle's energy loss rate 
is obtained as 
\begin{eqnarray}
\gamma&=&\left({t\over t_0}\right)^{-\alpha_V}\,{\gamma_0\over1+\gamma_0\psi(t,t_0)},
\label{eq:gam1}
\\
\psi(t,t_0)&=&\int^{t}_{t_0}
\left({t'\over t_0}\right)^{\alpha_{_V}}{dt'\over\tau_1},
  \label{eq:chi1}
\end{eqnarray}
where one assumes that the volume expands as $V=V_0(t/t_0)^{3\alpha_V}$ with 
$V_0$ the initial volume at time $t_0$. The integration in (\ref{eq:chi1}) can be carried 
out provided that a specific model for evolution of magnetic fields in the lobes is given.
If magnetic fields are completely tangled, they can be treated as fluids. When the magnetic fields
are in equipartition with particles, the density of magnetic energy decreases with time,
written as $U_B=U_{B0}(t/t_0)^{-\alpha_B}$, 
where $U_{B0}$ is the initial density at $t_0$ and the index $\alpha_B$ is a model-dependent
constant. A summary of $\alpha_B$ and $\alpha_V$ is given in Table~\ref{tab:alpha}.
The expansion law for magnetic fields is obtained by assuming equipartition with
particles. In the limiting-pressure model, $\alpha_B$ is obtained using (\ref{eq:Dt3}).
Eq (\ref{eq:chi1}) can be expressed as
\begin{eqnarray}
\psi(t,t_0)&=&t_0
\Biggl\{
{1\over(1-\alpha')\tau_{s0}}\left[
\left({t\over t_0}\right)^{1-\alpha'}-1\right]
\nonumber\\
&&
+{1\over(1-\alpha_{_V})\tau_{_{ICS}}}\left[
\left({t\over t_0}\right)^{1-\alpha_{_V}}-1\right]\Biggr\},
\label{eq:chi2}
\end{eqnarray}
where $\tau_{s0}=3m_ec^2/(4\sigma_TcU_{B0})$ 
and $\alpha'\equiv\alpha_B+\alpha_V$.

\begin{deluxetable}{llll}
\tabletypesize{\scriptsize}
\tablecaption{Expanson laws \label{tab:alpha}}
\tablewidth{0pt}
\tablehead{
\colhead{Source}&\colhead{Model} & \colhead{$\alpha_B$}   & \colhead{$\alpha_V$}}
\startdata
FR I & PLM & $\beta/(3-\beta)$ &     $1/3$ \\
FRII  & FEM  & $(4+\beta)/(5-\beta)$ &     $(4+\beta)/\Gamma(5-\beta)$ 
\enddata
\tablenotetext{}{PLM--the limiting pressure model. FEM--the free-expansion model.
 The constant-pressure model can be
regarded as a special case of PLM with $\beta=0$. The parameters
$\beta$ and $\Gamma$ are given in Table~\ref{tab:parameters}}
\end{deluxetable}

\subsection{Characteristic ages}

The evolution of the emitting plasma in the lobes is characterized by three time scales:
the adiabatic loss time, denoted by $\tau_a$, the synchrotron time
$\tau_s/\gamma$  and the ICS time $\tau_{_{ICS}}/\gamma$, where
$\tau_s$ and $\tau_{_{ICS}}$ are defined in (\ref{eq:b}).
For ICS, one only considers the CMB radiation as the seed photons.
The three time scales determine relative importance of these three
energy loss processes in the evolution.

In the early phase both adiabatic and synchrotron losses 
can be important. Setting $\tau_a=\tau_s/\gamma$, one obtains the 
characteristic age that separates the two energy loss regime,
estimated as
\begin{eqnarray}
t_a=t_0\left({\alpha_V\tau_{s0}\over t_0}\right)^{4/(4-3\alpha_B)}
\left({3\nu_{B0}\over4\nu}\right)^{2/(4-3\alpha_B)},
\label{eq:ta} \end{eqnarray} where the Lorentz factor is assumed to be
$\gamma_*=(4\nu/3\nu_{B0})^{1/2}(t/t_0)^{\alpha_B/4}$.
For $\alpha_B>4/3$, the synchrotron losses dominate initially and
at $t>t_a$ the adiabatic losses overtake the synchrotron losses.
For $\alpha_B<4/3$, the adiabtaic losses dominate first and
then the synchrotron losses become important at $t>t_a$.

One may compare the ICS losses with the synchrotron losses.
The energy loss rate due to ICS is $\gamma/\tau_{_{ICS}}\sim
4\gamma_*\sigma_TcU_{_{CMB}}/(3m_ec^2)$,
where $U_{_{CMB}}=3.6\times10^{-14}(1+z)^4\,{\rm J}\,{\rm m}^{-3}$ is
the CMB energy density at a redshift $z$.
One may express $U_{_{CMB}}$ in terms of an effective magnetic field $B_{CMB}\approx
3.2\times10^{-10}(1+z)^2\,\rm T$.
Equating the ICS energy loss rate to the synchrotron loss rate
yields a characteristic age:
\begin{eqnarray}
t_b=t_0\left({B_0\over B_{CMB}}\right)^{2/\alpha_B}.
\label{eq:tb}
\end{eqnarray}
The ICS losses become dominant over the synchrotron losses at $t>t_b$.

The characteristic age at which the adiabatic phase
switches to the ICS phase is estimated to be
\begin{eqnarray}
t_c=t_0\left({\alpha_{_V}\tau_{_{ICS}}\over t_0}\right)^{4/(4+\alpha_B)}
\left({3\nu_{B0}\over4\nu}\right)^{2/(4+\alpha_B)}.
\label{eq:tc}
\end{eqnarray}
Similar to the derivation of (\ref{eq:ta}), the derivation of (\ref{eq:tc})
involves replacing $\gamma$ by $\gamma_*$.  

\subsection{Analytical solutions}

We assume that particles are injected at a
constant rate with spectrum $q_l=q_0\gamma^{-p}$ and the initial 
condition $N(\gamma,t_0)=0$. The formal solution for (\ref{eq:CEq}) 
is written down in Appendix. There are two limits in which the exact analytical forms 
are well known. The first limit is when the adiabatic losses are dominant.
One may set $\tau_1\to\infty$, which leads to the exact solution \citep{es89},
\begin{eqnarray}
N(\gamma,t)={q_0\gamma^{-p}t\over1+(p-1)\alpha_{_V}}
\Biggl[
1-\left({t_0\over t}\right)^{(p-1)\alpha_{_V}+1}
\Biggr].
\label{eq:Nad}
\end{eqnarray}
Since the second term in the square brackets is generally much smaller than 1,
Eq (\ref{eq:Nad}) is approximately $\propto\gamma^{-p}$, which 
implies that the spectral slope is not affected by the adiabatic losses and
the whole spectrum raises proportionally with time.

The second limit is when the energy losses are due to ICS 
of the CMB radiation or due to synchrotron losses in constant magnetic fields. 
One has the well-known form \citep{k62,m80}: 
\begin{eqnarray}
N(\gamma,t)&=&{\tau_0q_0\gamma^{-p-1}\over p-1}
\Biggl[\!1\!-\!
\biggl(1-{(t-t_0)\gamma\over\tau_0}\biggr)^{p-1}\Biggr],
\label{eq:NSC}
\end{eqnarray}
where $\tau_0=\tau_{_{ICS}}$ for the ICS losses and $\tau_0=\tau_s$ for the 
synchrotron losses. Eq (\ref{eq:NSC}) must be subject to the condition $t-t_0\leq\tau_0/\gamma$.
When $\gamma\ll \tau_0/(t-t_0)$, i.e. the cooling time is much longer than 
the age, one has $N(\gamma,t)\approx q_0(t-t_0)\gamma^{-p}$.
This low energy limit can easily be understood from the continuity equation in which
the time derivative term, which describes the temporal evolution of the particle spectrum,
is more important than the convection (in $\gamma$) term that corresponds to 
radiative cooling of the emitting particles. Thus, the particle spectrum 
is $\propto t$ and its shape do not change. 
In the opposite limit in which the cooling time is much shorter than the age,
$\gamma>\tau_0/(t-t_0)$, the convection term in (\ref{eq:CEq}) 
is more important than the time derivative term. One can directly integrate 
(\ref{eq:CEq}) over $\gamma$ to obtain $N(\gamma,t)=q_0\tau_0\gamma^{-p-1}/(p-1)$.
Since for radio sources, it is usually true that the radiative cooling time
is considerably shorter than the source age, the particle spectrum in
the high energy approximation is the more relevant.

Apart from these two special cases, a third case of relevance, especially for
low-luminosity sources (cf. Sec 5), is the synchrotron losses in magnetic fields that slowly
decay with time. There is no simple analytical solution, though the particular case
where $\alpha_B<1$, $p=2$ and 3 was discussed in \citet{es89}. 
However, one can express the formal solution in terms of the hypergeometric 
function. The derivation is outlined in Appendix.
Here we only discuss the case $\alpha_B=1$, which corresponds to 
$\beta=3/2$ in the pressure-limiting expansion (cf. Sec. 5). Other examples are 
considered in the Appendix. The solution for $\alpha_B=1$ has the form,
\begin{eqnarray}
N(\gamma,t)&=&
{q_0t\gamma^{-p} e^{-1/\xi}\over (p-1)\xi}\nonumber\\
&&\times\Biggl[
M\left(p-1,p,1/\xi\right)-\left(1-\xi\ln(t/t_0)\right)^{p-1}
\nonumber\\
&&\times
M\left(p-1,p,1/\xi-\ln(t/t_0)\right)\Biggr],
\label{eq:NSyn}
\end{eqnarray}
where $\xi=\gamma t_0/\tau_{s0}\leq 1/\ln(t/t_0)$ and $M(a,b,x)$ is the Kummer's function
which has the asymptotic properties $M(p-1,p,x)\approx (p-1)x^{-1}e^x$ for $x\gg1$
\citep{as65}.  Thus, in the low energy limit $\xi\ln(t/t_0)\ll1$, one has 
\begin{equation}
N(\gamma,t)\approx q_0t\gamma^{-p}\Biggl[1-{t_0\over t}\biggl(1-{\gamma
t_0\over\tau_{s0}}\ln\left({t\over t_0}\right)\biggr)^{p-2}\Biggr], 
\label{eq:NSynL}
\end{equation}
which is similar to the low energy limit of (\ref{eq:NSC}).
In the high-energy regime, one has (cf. Appendix) 
\begin{equation}
N(\gamma,t)\approx {q_0\tau_{s0}\over p-1}\gamma^{-p-1}M(p-1,p,1/\xi).
\label{eq:NSynH}
\end{equation}
The spectrum with an initial power-index $p$ steepens to $\sim p+1$.

\section{High-luminosity radio galaxies}

Since FR IIs have already been considered in the literature, here
we rederive the main features of their evolutionary tracks for
nearby high luminosity sources ($z\ll1$), in particular the `knee' feature due to 
the transition from the expansion dominated by adiabatic 
losses to that by ICS of the CMB. To model the radiative evolution of 
the lobe we ignore the details of how relativistic particles are injected
into the lobe. One model for the injection is that the particles are
accelerated in the shock at the hot spots. The accelerated particles diffuse 
across the head region where they are subject to synchrotron losses. Since the 
pressure in the head region remains approximately constant as suggested 
from observations, the accelerated particles are
subject to severe adiabatic losses as they enter the lobe whose pressure is a decreasing
function of time \citep{betal99}. As a result, the radio power declines more
rapidly than that inferred from observations. One possible remedy for this enhanced 
loss is that particle re-acceleration is ongoing in the lobe \citep{mk02}.   
This would lead to similar results to that obtained by \citet{kb07} based on the 
time-independent injection. Here without going into a specific injection model, we 
adopt the similar assumption that the injection is time independent, with 
a power-law energy distribution.

\begin{figure}
\includegraphics[width=7cm]{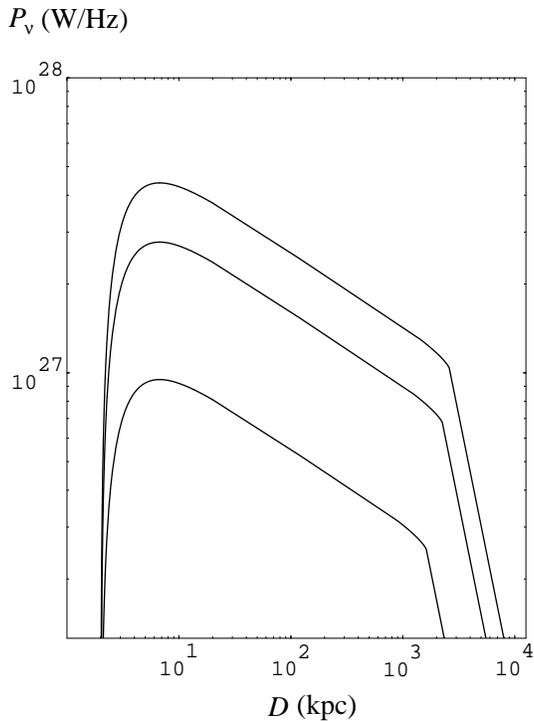}
\caption{$P_\nu$--$D$ tracks for FR IIs at $1.4\, \rm GHz$.
The plots from top to bottom correspond to $Q_j=2\times10^{40}\,\rm W$, $5\times10^{39}\,\rm W$
and $10^{39}\,\rm W$, respectively. The initial size and core radius are assumed to
be $2.5\,\rm kpc$.}
\label{fig:PD1}
\end{figure}

The evolution of the emitting plasma in the lobes 
can be characterized by three separate phases: 1) the initial
build-up phase in which the total energy in the lobe increases with time due to 
the injection of both kinetic energy by relativistic particles and magnetic energy, 
2) the adiabatic phase, and 3) the ICS phase.
The relevant times $t_a$, $t_b$ and $t_c$ can be estimated in a specific 
model for the lobe pressure $p_l$. We follow the procedure in 
\citet{betal99} by setting the pressure to that downstream of the bow
shock, giving $p_l\propto t^{-(4+\beta)/(5-\beta)}$. We also ignore the large-scale 
ordered magnetic fields and set the magnetic pressure to
$p_l$. The adiabatic expansion $p_lV^{\Gamma}={\rm const}$ gives 
$\alpha_V=(4+\beta)/\Gamma(5-\beta)$ \citep{ketal97}. One has
\begin{equation}
t_a\approx
6.1\times10^{-3}\left({t_0\over0.1\,{\rm Myr}}\right)^{33/5}B^{42/5}_{50}\nu^{14/5}_1\,{\rm Myr},
\label{eq:ta2}
\end{equation}
\begin{equation}
t_b\approx 62\left({t_0\over0.1\,{\rm Myr}}\right)B^{14/11}_{50}(1+z)^{-28/11}\,{\rm Myr},
\end{equation}
\begin{equation}
t_c\approx 135\left({t_0\over0.1\,{\rm Myr}}\right)^{0.28}B^{0.36}_{50}\nu^{-0.36}_1(1+z)^{-1.4}\,
{\rm Myr},
\end{equation}
where one assumes that $B_{50}=B_0/(50\,{\rm nT})$, $\nu_1=\nu/(1\,\rm GHz)$
and $\beta=3/2$. That $t_b\gg t_a$ implies that synchrotron losses are dominant only in the 
very early, build-up phase of the evolution.

For high-power jets, by substituting (\ref{eq:tc}) for (\ref{eq:Dt2}), one 
estimates the characteristic size at which the ICS losses
dominate. This corresponds to
a break or `knee' at
\begin{equation}
D_c\approx D_0\left({t_c\over t_0}\right)^{3/(5-\beta)}.
\end{equation}
Since in the adiabatic regime, the particle spectrum remains the same as 
the injection spectrum, the total spectral power for $D<D_c$ is $P_\nu\propto U^{(1+p)/4}_Bt$, 
where the power-law index of the electron distribution, $p$, is assumed to 
be constant.  Since $t\propto D^{(5-\beta)/3}$,
the $P_\nu$--$D$ relation is derived to be
\begin{equation}
P_\nu\propto D^{-\delta},\quad\quad\delta={\textstyle{1\over12}}
(5-\beta)\left[(1+p)\alpha_B-4\right].
\label{eq:PD1}
\end{equation}
Eq (\ref{eq:PD1}) reproduces (3) in \citet{kb07} 
when $p=2$. One obtains $\delta=(7\beta-8)/12\approx5/24$ for $\beta=3/2$.

When energy loss due to ICS is important, the number
of particles with large Lorentz factors decreases leading to steepening of 
the particle distribution with the index increasing to $p+1$. The particle 
spectrum becomes time-independent, which is in contrast to the low-energy regime
or the adiabatic regime in which the particle spectrum increases with time.
In the ICS regime ($D>D_c$), the total power is $P_\nu\propto U^{(2+p)/4}_B\sim D^{-\delta}$, 
with an index
\begin{equation}
\delta={\textstyle{1\over12}}(5-\beta)(2+p)\alpha_B\approx 1.8.
\end{equation}
The index is similar to that found in \citet{kb07} (their Eq 7).
The approximation is obtained when $p=2$ and $\beta=3/2$. As a result, the
$P_\nu$--$D$ power-law in this regime is significantly steeper than (\ref{eq:PD1}).

Figure~\ref{fig:PD1} shows three $P_\nu$--$D$ tracks at $1.4\,\rm GHz$, with three 
different input powers.  Here both the initial size and the core radius are taken to be 2 kpc. 
The tracks are obtained using the analytical solutions (\ref{eq:Nad}) and (\ref{eq:NSC})
with $\theta=0.5$, $\eta=0.5$, $\rho_c=1.7\times10^{-22}\,{\rm kg}\,{\rm m}^{-3}$,
$\beta=3/2$, and $\Gamma=4/3$ (which is appropriate for relativistic plasmas).
We assume the particle spectrum to be the typical one from the standard diffusive shock 
acceleration, characterised by a power-law with an index $p=2.1$ and the lower- and
upper-cutoff, $(\gamma_1,\gamma_m)=(5,10^7)$ (cf. Eq [\ref{eq:ql}]).
The magnetic field  at the hot spots is assumed to be $B_0=30\,\rm nT$ \citep{betal99}.
Since the initial particle spectrum is assumed to be zero, the spectral power, $P_\nu$,
increases rapidly to the phase when the spectral power reaches the maximum and
the adiabatic losses set in. During the initial phase, the synchrotron losses
compete against replenishing of new particles from the injection, with the latter
dominating. The radio power decreases slowly as the source size grows, primarily
due to energy losses through adiabatic expansion; during this phase, the power
spectrum is $P_\nu\sim \nu^{-\alpha}$ with $\alpha=(p-1)/2$.
When the size exceeds the characteristic size $D_c$, the power starts 
to decrease rapidly due to the ICS losses. Since $D_c\propto D^{0.28}_0$ with $\beta=3/2$, 
the characteristic size $D_c$ is not particularly sensitive
to the initial size $D_0$. As shown in the figure, the location $D_c$ of the 
break, i.e. the `knee' feature  remains roughly the same
for different $D_0$. In the ICS regime, the power spectrum steepens to
$\alpha=p/2$. The tracks are the most strongly affected by the input power
$\eta Q_j$, the magnetic field at the hot spot, $B_0$, and the index for the density profile,
$\beta$. The track height increases if one increases the input power or the magnetic field 
at the hot spot or both. In general, a larger $\beta$ leads to steepening of the track slope in
both the adiabatic regime and ICS regime.

\begin{deluxetable*}{llll}
\tabletypesize{\scriptsize}
\tablecaption{Model parameters for low-luminosity sources \label{tab:parameters}}
\tablewidth{0pt}
\tablehead{
\colhead{Component}&\colhead{Parameters}   &  \colhead{Definition}  &\colhead{Typical values}  
}
\startdata
Jet& $Q_j$      &Injection power (W)   & $10^{32}-5\times10^{36}$  \\
& $\eta$      &Efficiency of power injection    & $0.5$  \\
Lobe &$\chi$     & Geometric factor  & $1$\\
& $\Gamma$     & Adiabatic index  & $4/3$\\
& $p$      &Relativistic particle spectral index   & $2.1$ \\
& $\gamma_1$ & Minimum Lorentz factor & $5$\\
& $\gamma_m$ & Maximum Lorentz factor & $10^7$\\
Environment& $\rho_c$      &Core density (${\rm kg}\,{\rm m}^{-3}$)  & $1.7\times10^{-23}$ \\
& $r_c$      &Core radius (kpc)  & $1.5-3$ \\
& $\beta$    &Power index of the density radial profile                    &
$1.5$ (PLM), $0$ (CPM${}^*$)             \\
& $T_0$      &Temperature (K) & $10^7$ \\
& $p_c$      &Core pressure (Pa) & $3\times10^{-11}$ 
\enddata
\tablenotetext{}{${}^*$ CPM--the constant pressure model.
Since $p_c=\rho_ck_BT_0/m_p$, the pressure $p_c$ is treated as an independent
parameter in our model.}
\end{deluxetable*}

\section{Low-luminosity radio galaxies}

The pressure-limiting expansion model is applied to FR I sources. A
low-power jet may undergo free expansion initially, but such initial phase 
can only last very briefly. Therefore, one can ignore this 
phase and assume that the jet expands slowly in the pressure-confined environment. 

\subsection{Observational data}

To test the models, we assembled a complete sample of low-luminosity radio galaxies
from the northern zone of 2dF Galaxy Redshift Survey (2dFGRS) \citep{cetal01}
which has 1.4\,GHz radio continuum data available from both the NVSS
\citep{cetal98} and FIRST \citep{betal95} surveys.
The 2dFGRS radio sample was selected
by matching the NVSS and 2dFGRS catalogues using the techniques described
by  \citet{setal02}.  We then restricted our final sample to galaxies
which satisfied the following criteria:
\begin{itemize}
\item
Classified as an AGN on the basis of the 2dFGRS optical spectrum (i.e. star--forming
galaxies are excluded).
\item
The total NVSS 1.4\,GHz flux density is at least 10\,mJy, to allow an accurate
measurement of the radio-source angular size.
\item
The galaxy is also detected in the FIRST catalogue, which has higher angular
resolution (with a 5\,arcsec beam, compared to 45\,arcsec for NVSS).
\end{itemize}

\subsection{Radio luminosity and source size}

This selection produced a final sample of 375 low-luminosity radio galaxies,
with redshifts in the range $z=0.02$ to 0.3 (median $z=0.136$). All 375 sources
were spatially resolved in the FIRST survey, so there are no upper limits on
the angular size of these radio sources.

We calculated the  1.4\,GHz radio power and largest linear size (LLS) of each
source from the measured NVSS flux density and the largest angular size (LAS)
measured  from either the NVSS catalogue or (if unresolved in NVSS) the
FIRST catalogue. We adopted a cosmology with 
${\rm H}_0=71\,{\rm km}\,{\rm s}^{-1}\,{\rm Mpc}^{-1}$,
$\Omega_{\rm m}=0.23$ and $\Omega_{\rm \Lambda}=0.73$ for these calculations.
In the few cases where a source was resolved into two or more
components in NVSS or FIRST, we measured the LAS across all components.
The 1.4\,GHz radio power of galaxies in this sample spans a range from
$10^{22}$ to $10^{26.5}\,{\rm W}\,{\rm Hz}^{-1}$ (median $10^{24.2}\,{\rm W}\,{\rm Hz}^{-1}$),
and the LLS values range from 0.4\,kpc to 715\,kpc (nedian 51\,kpc).

\begin{figure}
\includegraphics[width=7cm]{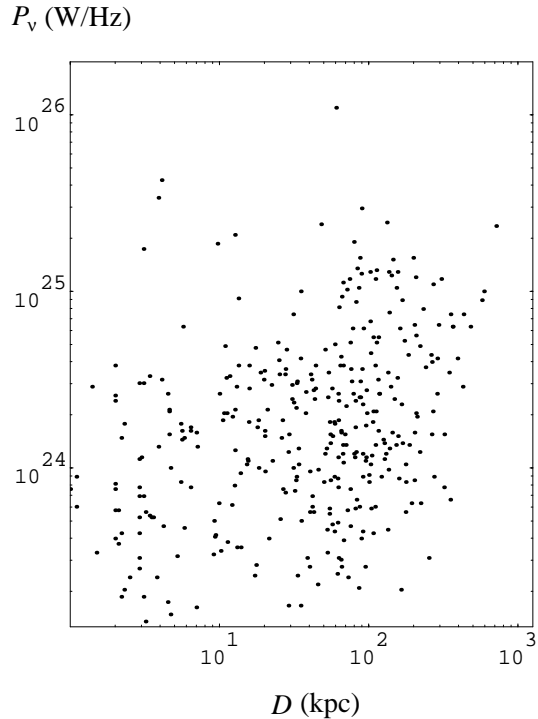}
\caption{Local low-luminosity radio galaxies at $1.4\,\rm GHz$, from 
the 2dFGRS sample described by \citet{setal02}.
}
\label{fig:LRG}
\end{figure}

The radio $P_\nu$-$D$ diagram for the sample is shown in Figure \ref{fig:LRG}.
There are two interesting features
that can be used to constrain the modeling. First, there
are fewer sources with a small size in the high luminosity region.
Second, there is a deficit in the number of sources with large sizes in the low luminosity
region. The second feature may be understood qualitatively as due to the flux limiting effect
(cf. Sec. 2.5), i.e. because of the lower limit on the observable flux,
the $P_\nu$--$D$ have a cutoff $P_\nu>F_{\nu*}D^2$. 
The first feature suggests that the source, as we argue, expands at a rate considerably
faster than the constant-pressure expansion (see Sec 5.2). As a result of increased 
losses, the initial increase in the luminosity is more gradual than that 
predicted by the constant-pressure expansion model.

\begin{figure}
\includegraphics[width=7cm]{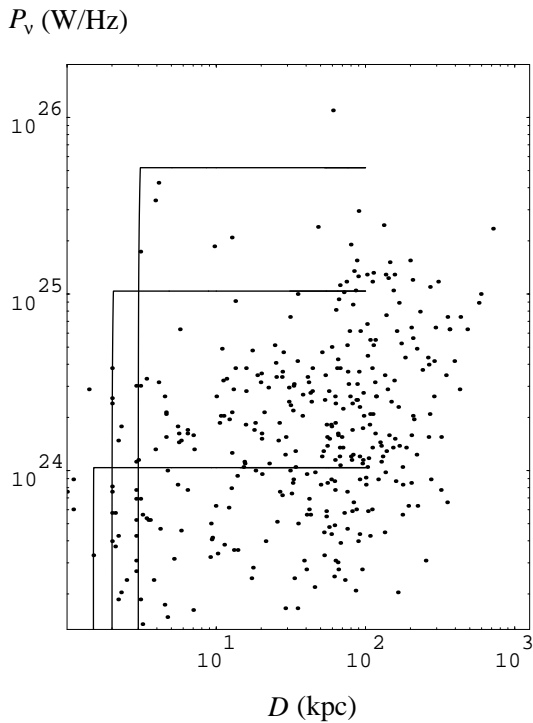}
\caption{$P_\nu$--$D$ tracks for low-luminosity radio galaxies.
The tracks are obtained in the constant-pressure expansion model.
Plots from top to bottom correspond to
$(Q_j, D_0)=(5\times10^{36}\,{\rm W}, 3\,{\rm kpc})$,
$(10^{36}\,{\rm W}, 2\,{\rm kpc})$,
$(10^{35}\,{\rm W}, 1.5\,{\rm kpc})$, respectively
The tracks increase much more slowly in the initial phase compared with
FR IIs. The plots are terminated at $t=10^4\,\rm Myr$.
}
\label{fig:PD2}
\end{figure}

\subsection{Power-size tracks}

We consider the scenario of constant-pressure expansion first. The
external pressure $p_{ex}$ is held constant with $p_l\sim p_{ex}$. We show that 
the constant-pressure model would generally underpredict large-size sources.
The model overpredicts small-size sources with high luminosities.
If equipartition applies, one has $p_B\sim p_l\sim {\rm const}$. 
Setting $\alpha_V=1/3$ and $\alpha_B=0$, one has 
\begin{equation}
t_a={\textstyle{1\over3}}\tau_{s0}\left({3\nu_{B0}\over4\nu}\right)^{1/2}
\approx 0.4B^{-3/2}_{10}\nu^{-1/2}_1\,{\rm Myr}.
\label{eq:ta3}
\end{equation} 
Since $t_b\to\infty$, the synchrotron losses are always dominant over the
ICS losses. The synchrotron losses dominate over the adiabatic losses
at $t>t_a$. Eq (\ref{eq:ta3}) implies the characteristic size
\begin{eqnarray}
D_a&=&D_0\left({\tau_{s0}\over3 t_0}\right)^{1/3}
\left({3\nu_{B0}\over4\nu}\right)^{1/6}\nonumber\\
&\approx& 1.6D_0B^{-1/2}_{10}\nu^{-1/6}_1\left({t_0\over0.1\,{\rm Myr}}\right)^{-1/3}.
\end{eqnarray}  
One may express the total power as a power-law of the size,
$P_\nu\sim D^{-\delta}$. For $D<D_a$, one has $\delta=-3$. Since the particle spectrum 
in the synchrotron regime with $t\ll t_0+\tau_{s0}/\gamma_*$ has the same slope as 
in the adiabatic case, one has $\delta=-3$ for $D>D_a$ as well.
The particle spectrum due to the synchrotron losses in constant magnetic fields
has the same form as (\ref{eq:NSC}) with $\tau_{ICS}$ replaced by
$\tau_{s0}$.
The power grows with increasing size. However, when $t\sim t_0+\tau_{s0}/\gamma_*$,
corresponding to the size $\sim D_0[1+(\tau_{s0}/t_0)(3\nu_{B0}/4\nu)^{1/2}]^{1/2}$,
the spectrum of the emitting particles steepens from
$p$ to $p+1$ and becomes independent of time. As a result, the power remains 
constant with $\delta=0$. The model would predict a steep spectrum with
$\alpha=p/2=1$ for $p=2$. Although there is no direct measurements 
of the radio spectral index distrbution in the sample of 2dFGRS
galaxies used to derive the $P_\nu$--$D$ diagram in Figure~\ref{fig:LRG},
there is strong evidence from the lower frequency data that low-luminosity 
radio galalxies on average have flatter radio spectra \citep{metal03,betal99}. 
A possible interpretation for this spectral trend is
that for low-luminosity radio galaxies, the core component becomes important
\citep{setal94}, i.e. one may obtain flat spectra by appealing to self-absorption.
However, the steep spectral feature predicted by the constant-pressure model would extend 
to the majority of low-luminosity sources, which is not consistent with observations.  

Figure \ref{fig:PD2} shows plots of $P_\nu$ as a function of $D$ for 
low-luminosity sources (FR Is), obtained in the constant-pressure model with
$z=0$ and $\beta=0$. The model parameters and their typical values are listed in 
Table~\ref{tab:parameters}. If one considers the external medium as an ideal gas,
the three parameters, $\rho_c$, $T_0$ and $p_c$ are not indepedent. Here we 
treat the external pressure as an independent parameter and is assumed to be 
$p_{ex}=p_c=3\times10^{-11}\,\rm Pa$. As we assume that the magnetic energy is in 
equipartition with kinetic energy of particles, the magnetic field pressure is not an
independent parameter. The justifications for the choice of the 
particle spectrum are the same as that for high-luminosity sources--particles
are accelerated by diffusive shocks (cf. Sec. 4). The typical value of the
adiabatic index, $\Gamma=4/3$, is applicable for relativistic plasmas.
(For a cold plasma, one has $\Gamma=5/3$.) 
The dividing line between high- and low-luminosity sources in terms of
the input power from the jet is $\sim 10^{37}\,\rm W$ \citep{lo96}. The plots are
obtained using three different input powers that are below this characteristic 
power. The dots represent the observations (cf. Sec. 5.2). The plots are terminated 
at $t=10^4\,\rm Myr$. The luminosities remain 
constant after reaching their maxima. The luminosities
may eventually drop below detection. For example, one may invoke a model in which the 
lobes are assumed to expand into a much lower density region, which
can result in a rapid decrease in the luminosity \citep{gw88}. However, as 
shown in figure~\ref{fig:PD2}, the constant-pressure model underpredicts the 
large-size sources while predicts excess number of the small-size sources.

\subsection{Pressure-limiting approximation}

\begin{figure}
\includegraphics[width=7cm]{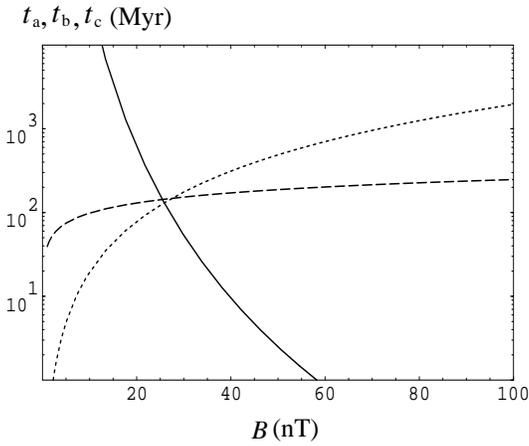}
\caption{Plots of $t_a$ (solid), $t_b$ (dotted), $t_c$ (dashed) with
$t_0=0.02\,\rm Myr$, $z=0$ and $\beta=3/2$.
}
\label{fig:ct}
\end{figure}

When the lobe expansion is pressure limited with $\beta\neq0$, the three characteristic
times discussed in Sec 3.2 are now given by
\begin{equation}
t_a\approx
2\times10^{-2}\left({t_0\over0.1\,{\rm Myr}}\right)^{-3}B^{-6}_{50}\nu^{-2}_1\,{\rm Myr},
\label{eq:ta4}
\end{equation}
\begin{equation}
t_b\approx 2.4\times10^3\left({t_0\over0.1\,{\rm Myr}}\right)B^2_{50}(1+z)^{-4}\,{\rm Myr},
\end{equation}
\begin{equation}
t_c\approx 2.6\times10^2\left({t_0\over0.1\,{\rm
Myr}}\right)^{1/5}B^{3/7}_{50}\nu^{-3/7}_1(1+z)^{-1.6}\,
{\rm Myr},
\end{equation}
where $\beta=3/2$ and $\alpha_B=\beta/(3-\beta)$. Here the estimate for 
$\alpha_B$ is obtained assuming energy equipartition between 
the magnetic field and particles (cf. Table \ref{tab:alpha}). 
Plots of $(t_a, t_b, t_c)$ are shown in figure~\ref{fig:ct}. 
Like the high-luminosity sources,
both $t_a$ and $t_b\propto B^2_0$ are strongly dependent on the initial magnetic field
which is assumed to be the magnetic field in the flare region.
At $B_0>26\,\rm nT$, the energy loss processes dominate along the evolutionary track 
in the following order: adiabatic, synchrotron, ICS; in the low field case $B_0\leq
26\,\rm nT$, the ICS dominance occurs before the synchroron loss. 
It is interesting to note that $t_b$ is very sensitive to $z$ 
compared to high-luminosity sources. For local sources ($z\ll1$), $t_b$ is 
over $100\,\rm Myr$.  Thus, for local sources, the ICS losses play a role only in 
the very late stage of the evolution.

\begin{figure}
\includegraphics[width=7cm]{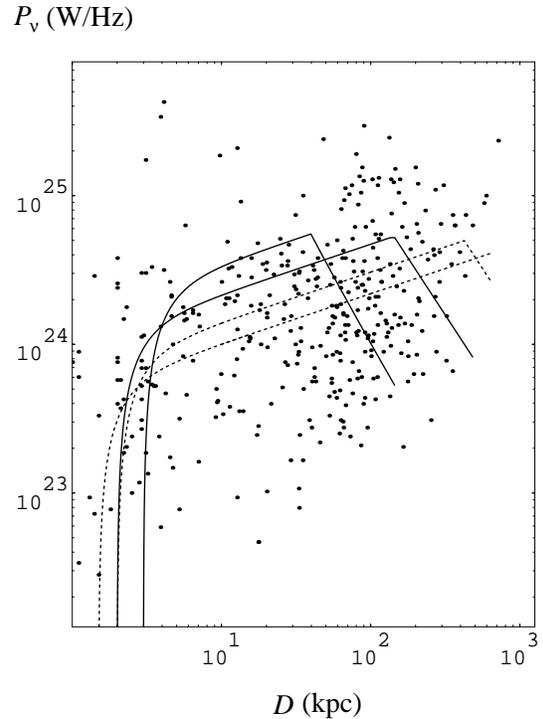}
\caption{As in figure~\ref{fig:PD2} but obtained in the
pressure-limiting model with $Q_j=1.5\times10^{36}\,\rm W$ and 
$\beta=3/2$. The solid and dashed lines correspond to $r_c=2\,\rm kpc$ 
and $r_c=1.5\,\rm kpc$, respectively.
}
\label{fig:PD3}
\end{figure}

\begin{figure}
\includegraphics[width=7cm]{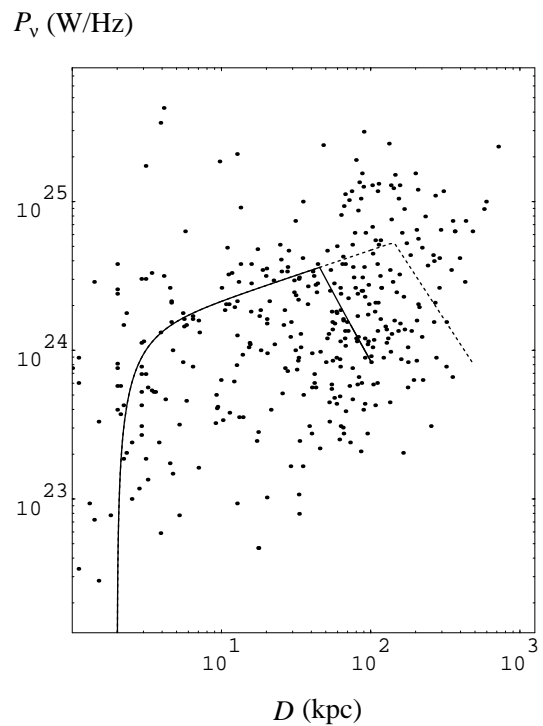}
\caption{As in figure~\ref{fig:PD3} for two different jet powers,
$Q_j=10^{36}\,\rm W$ (solid line) and $1.5\times10^{36}\,\rm W$
(dashed line). We assume $D_0=2\,\rm kpc$.
}
\label{fig:PD4}
\end{figure}

The $P_\nu$--$D$ tracks can be obtained analytically. There are three relevant phases:
the initial rapid rise, followed by the more gradual increase in luminosity,
and then decline in luminosity.
In the gradual increasing phase, the particle spectrum is $N(\gamma,t)\propto t\gamma^{-p}$.
The $P_\nu$--$D$ track can be approximated by a power-law,
$P_\nu\sim D^{-\delta}$. From (\ref{eq:Pnu}) using (\ref{eq:Dt3}), one finds $\delta=-(3-\beta)
[4-\alpha_B(1+p)]/4=-3/8$ for $p=2$ and $\beta=3/2$. When the self-absorption effect is neglected,
the corresponding radio spectrum is $\alpha=(p-1)/2$.
When $t\sim t_a$, the source enters the declining phase in which the magnetic fields in
the lobe become too low and given a fixed observation frequency, the Lorentz factor of
the emitting particles shifts to a much higher value. The time $t_a$ when the track
starts to turn over is sensitive to the initial time $t_0$ and magnetic field $B_0$.
For $t_0=10^4\,\rm yr$ and $B_0=50\,\rm nT$, one estimates $t_a\approx 20\,\rm Myr$.
For $t>t_a$, the synchrotron losses become dominant and the particle spectrum
is stationary with the spectral slope steepening to $p+\Delta$ with $\Delta$
a function of $\beta$. This leads to $\delta=(p+\Delta)\beta/2$. The radio
spectrum steepens to $\alpha=(p+\Delta-1)/2$. For $\beta=3/2$,
one has $\Delta\sim 1$.

\begin{figure}
\includegraphics[width=7cm]{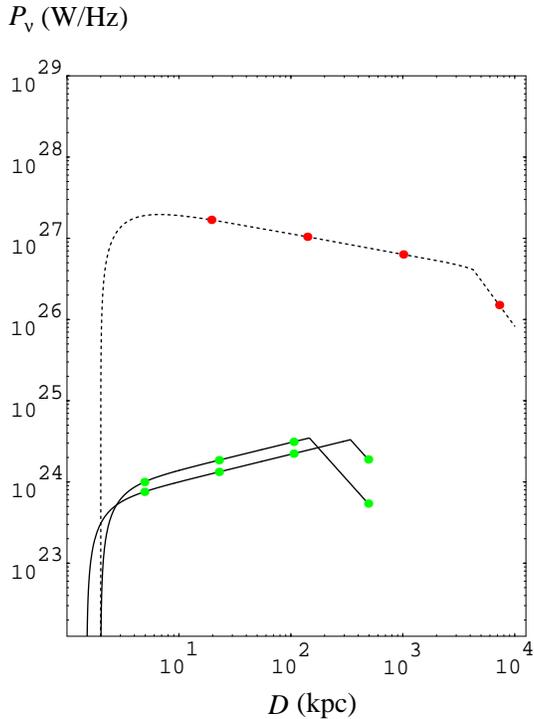}
\caption{High-luminosity sources (dashed) and low-luminosity sources (solid).
The dots (from left to right) on each curve indicate ages; for high-luminosity
sources, they represent $0.1\,\rm Myr$, $1\,\rm Myr$, $10\,\rm Myr$ and $10^2\,\rm Myr$.
For low-luminosity sources, the dots represent $1\,\rm Myr$, $10\,\rm Myr$,
$10^2\,\rm Myr$, and $10^3\,\rm Myr$. 
}
\label{fig:PD5}
\end{figure}

Figure~\ref{fig:PD3} shows the $P_\nu$--$D$ tracks in the pressure-limiting case 
with the parameters given in Table~\ref{tab:parameters}. Although there
is a spread in $\beta$, here we take the typical value $\beta=3/2$. 
The pressure at the core is assumed 
to be $p_c=3\times 10^{-11}\,\rm Pa$ corresponding the density $n_0=2\times10^5\,{\rm m}^{-3}$ 
and the temperature $T_0=10^7\,\rm K$. As in figure~\ref{fig:PD2}, the plots are 
overlaid on the observational data of local luminosity sources represented by dots. 
Figure~\ref{fig:PD4} shows how the turn-over depends on the input power $Q_j$.
For low $Q_j$, the track turns over at a much smaller size.
The evolution pace is characterized by $dD/dt\sim (D_0/t_0)(t/t_0)^{-(2-\beta)/(3-\beta)}$.
Since the initial size $D_0$ is treated as an input parameter in the plots, 
(\ref{eq:Dt3b}) implies that a small $Q_j$ corresponds to a long $t_0$, i.e. the source moves 
more slowly along the track than otherwise. Thus, the input power determines the evolution pace
along the track. As in the constant pressure case, we adopt equipartition in our calculation;
Thus, there is a direct connection between the magnetic pressure in the lobe and the external pressure. 
Since the synchrotron power is $\propto U^{(p+1)/4}_B$, the track height  
depends more strongly on $p_c$ than the jet input power ($Q_j$).
One can calculate $P_\nu(D)$ for an intermediate case $0<\beta<3/2$ as well. It can be shown 
that the resultant track increases rapidly in the initial phase which similar to the constant-pressure 
case but the radio power drops off rapidly at small sizes. This early steep drop off is due to
that the particle spectrum become very steep with $p+1/(1-\alpha_B)$ (cf. the comments below 
Eq [A10]). In the case of a steep density profile ($\beta>12/7$), 
one expects the track to change to the shape with a declining
trend. This can be understood qualitatively that the lobe expands more rapidly
in an external medium with rapid decreasing density in radial direction.

Figure~\ref{fig:PD5} shows comparison of FR Is (solid lines) and FR IIs (dashed line).
One assumes $Q_j=10^{40}\,\rm W$ and $(D_0,r_c)=(2\,{\rm kpc},2\,{\rm kpc})$
in plotting the dashed curve and  $Q_j=10^{36}\,\rm W$ and $(D_0,r_c)=(1.5\,{\rm kpc},1.5\,{\rm kpc})$
in plotting the upper solid curve and $(D_0,r_c)=(2\,{\rm kpc},2\,{\rm kpc})$ in the lower solid curve.  
Except for the initial brief increase, the luminosities of FR IIs decay
during their life time. By contrast, our model predicts that low-luminosity sources  
grow with increasing luminosities. The growth of radio power stops when the synchrotron 
or ICS losses become dominant over the adiabatic losses. For $Q_j=1.5\times10^{36}\,\rm W$, 
the radio power peaks at about $100\,\rm Myr$. 

\section{Conclusions and discussion}

We consider the evolution of low-luminosity radio galaxies with $z\ll1$. The radio 
power as a function of the source's size is derived based on a generic model 
that uses global parameters of the jet-lobe system. In the global model discussed
here, we assume that the lobe inflated by a low-power jet undergoes 
pressure-limiting expansion, in which the system is approximately in pressure-balance
against the external medium. Thus, the energy equation (Eq [\ref{eq:El}]) 
for the lobe can be solved by replacing the lobe pressure, $p_l$, with the external pressure,
$p_{\rm ex}$. The source's size as a function of time can be derived in both cases of  
constant external pressure and decreasing external pressure. In the calculation of the 
$P_\nu$--$D$ track we use the standard theory of nonstationary spectra of relativistic 
particles that are subject to both adiabatic and radiative losses. 
The predicted $P_\nu(D)$ can be compared with the $P_\nu$--$D$ diagram from observations.
The main conclusions are summarised as follows.

\noindent
1) The pressure-limiting expansion model (with $\beta\sim 3/2$) predicts $P_\nu$--$D$ 
tracks that are generally consistent with observations; in particular, it
predicts fewer small-size sources in the high luminosity region in the radio $P_\nu$-$D$ 
diagram. Since the emitting particles suffer expansion losses, the luminosity increases 
gradually untill the synchrotron or ICS losses become dominant.

\noindent
2) The constant-pressure expansion model (with $\beta=0$) generally 
predicts an overabundance of small-size FR I radio sources with a relatively 
high radio power and at the same time underpredicts the large-size FR I sources. 
These features are not consistent with observations (cf. Figure~\ref{fig:PD2}).

\noindent
3) In the pressure-limiting expansion model, the physical conditions
of the external medium, e.g.  $p_{\rm ex}$, play an important role in 
the evolution. By assuming equipartition, one expects to see a direct link
between the radio power and the external pressure, i.e. the higher 
external pressure the higher radio power. The $P_\nu$--$D$ track 
would also turn over at a much smaller size.

It is interesting to compare the model prediction for low--luminosity sources with 
that of high-luminosity sources. For low-power jets, the lobes expand slowly in near 
pressure equilibrium with the external pressure. Thus, FR Is evolve much more slowly 
compared with FR IIs. Our model predicts that the radio power of FR Is may increase 
throughout most of their life time. By contrast, the radio power of FR IIs declines in 
most of their life time (except for the brief initial rise) and can be well described 
by two power-laws. The initial rise for FR IIs is very brief, less than 0.1 Myr for 
the parameters adopted in Figure~\ref{fig:PD1}, while for FR Is, this can last more 
than 100 Myr. These differences suggest that these two types of source evolve differently.  

In our global model, both the details of the particle injection and
the effect of spatial diffusion are not treated. The effect of spatial 
diffusion can be important in particle transport--this is particularly the case
when particle acceleration is confined to a localized site, say the flare region (`hot spot') 
\citep{eetal97}. The accelerated particles need to diffuse across the region surrounding the
acceleration site. Note that this requirement may be relaxed if particles are accelerated
by multiple weak shocks distributed over an extended region in the lobe. 
Particle diffusion should strongly depend on plasma turbulence
in the region concerned. To obtain the solution for particle diffusion, one needs
to deal with the full diffusion-loss equation that includes particle diffusion
in plasma turbulence \citep{l94}.

One should emphasise that the pressure in the energy equation (\ref{eq:El}) is
the total pressure that may consist of radiating and nonradiating particles as well
as magnetic pressure. For convenience, we consider only one species of
particles (radiating particles). X-ray observations suggest that in some of the known FR I
sources, the equipartition pressure in the lobe is substantially lower than the external pressure
\citep{cetal08}. One possible explanation is that the missing pressure may
be provided by nonradiating particles due to entrainment. 
Although in principle, the case can be treated in a similar way to that presented
here for single species of particles by choosing a low $\eta$ and $\Gamma\to 5/3$,
An extension of this model to including multi-component plasmas in the lobe will 
be considered in future work.

Finally we comment on the equipartition assumption adopted in the calculation 
of the radio power. The magnetic fields are assumed to be 
completely entangled (i.e. we ignores the effect of the  mean field).
The equipartition assumption is reasonable as recent {\em Chandra} X-ray observations
suggest that the magnetic fields in lobes are close to the equipartition field \citep{betal08}.
It is worth noting that the frozen flux argument would suggest that the magnetic 
energy density in an expanding lobe decreases with time much faster than 
$\propto1/t^{\beta/(3-\beta)}$. To maintain equipartition, one requires 
either that the magnetic turbulence be generated in the lobe
\citep{d80} or that the magnetic energy be injected into the lobe \citep{es89}.

\acknowledgments
We thank Don Melrose for helpful comments on the manuscript. EMS thanks 
the Australian Research Council (ARC) for support.

\begin{appendix}

\section{Synchrotron losses in decaying magnetic fields}

Similar to the method discussed in \citet{es89}, one 
may solve (\ref{eq:CEq}) by solving the ordinary differential equation:
\begin{equation}
{dN(\gamma_0,t)\over dt}=b_\gamma(\gamma_0,t)N(\gamma_0,t)+q_l(\gamma_0,t), 
\label{eq:ODE}
\end{equation}
where $\gamma_0$ is treated a parameter and the functions $b_\gamma(\gamma_0,t)$,
$N(\gamma_0,t)$ and $q_l(\gamma_0,t)$ correspond to $\partial b/\partial\gamma$,
$N(\gamma,t)$ and $q_l(\gamma)$ with $\gamma$ written as 
a function of $(\gamma_0,t)$:
\begin{equation}
\gamma={\gamma_0\over1+\gamma_0\chi},\quad
\psi={t_0\over\tau_{s0}(1-\alpha_B)}\Biggl[\left({t\over t_0}\right)^{1-\alpha_B}-1\Biggr].
\label{eq:ER}
\end{equation}
The solution is
\begin{equation}
N(\gamma_0,t)=q_0t_0\gamma^{-p}_0
\Biggl[1+{\xi_0\over1-\alpha_B}\left(x^{1-\alpha_B}-1\right)\Biggr]^2I(\gamma_0,t),\quad
I(\gamma_0,t)\equiv
 \int^x_{1}\Biggl[1+{\xi_0\over1-\alpha_B}\left({x'}^{1-\alpha_B}-1\right)\Biggr]^{p-2}
dx',
\label{eq:NSynG}
\end{equation}
where $\xi_0=\gamma_0t_0/\tau_{s0}$ and $x=t/t_0$. 
The final form $N(\gamma,t)$ can be obtained by transforming (\ref{eq:NSynG})
 back to the domain $(\gamma,t)$, using
\begin{equation}
\gamma_0=\gamma\Biggl[1-{\xi\over1-\alpha_B}\left(x^{1-\alpha_B}-1\right)\Biggr]^{-1},
\quad
\xi_0=\xi\Biggl[1-{\xi\over1-\alpha_B}\left(x^{1-\alpha_B}-1\right)\Biggr]^{-1},
\quad \xi={\gamma t_0\over\tau_{s0}}.
\label{eq:transform}
\end{equation}

\subsection{Special cases}

Consider two special cases $\alpha_B=1/2$ and $\alpha_B=1$. For $\alpha_B=1$, one has
\begin{eqnarray}
N(\gamma_0,t_0)&=&q_0t_0\gamma^{-p}_0\left(1+\xi_0\ln x\right)^2I(\gamma_0,t), \label{eq:NSyn5}\\
I(\gamma_0,t)&=&\int^x_{1}\left(1+\xi_0\ln x'\right)^{p-2}
dx'\nonumber\\
&=&{e^{-1/\xi_0}\over \xi_0(p-1)}\Biggl[
(1+\xi_0\ln x)^{p-1}M(p-1,p,(1+\xi_0\ln x)/\xi_0)\nonumber\\
&&- M(p-1,p,1/\xi_0)\Biggr],
\label{eq:NSynSa}
\end{eqnarray} 
where $M(a,b,z)$ is the Kummer's function, defined as \citep{as65}
\begin{equation}
M(a,b,z)={\Gamma(b)\over\Gamma(b-a)\Gamma(a)}
\int^1_0e^{zt}t^{(a-1)}(1-t)^{b-a-1}dt.
\label{eq:KF}
\end{equation}
Eq (\ref{eq:NSyn}) can be obtained by substituting (\ref{eq:transform}) back to
(\ref{eq:NSyn5}) and (\ref{eq:NSynSa}).
The low energy limit (\ref{eq:NSynL}) can be obtained by taking the limit $\xi\ln(t/t_0)\ll1$. 
The high energy limit (\ref{eq:NSynH}) can be derived by taking either the limit
$\xi_0\ln (t/t_0)\to\infty$ in (\ref{eq:NSynG}) or the limit $\xi\ln(t/t_0)\to 1$
in the final form.

For $\alpha_B=1/2$, the integral can be evaluated exactly as
\begin{eqnarray}
I(\gamma_0,t)={1\over 2\xi^2_0p(p-1)}
\Biggl\{\biggl[1+2\xi_0(x^{1/2}-1)\biggr]^{p-1}\biggl[2\xi_0\Bigl(1+(p-1)x^{1/2}\Bigr)-1\biggr]
-(2\xi_0p-1)\Biggr\}.
\end{eqnarray}
This gives
\begin{equation}
N(\gamma,t)={q_0t_0\gamma^{-p}\over
2p(p-1)\xi^2}\Biggl\{
2\xi p\sqrt{x}-1-\biggl[1-2\xi(\sqrt{x}-1)\biggr]^{p-1}
\biggl[
2\xi(p-1+\sqrt{x})-1\biggr]\Biggr\}.
\label{eq:NSynSb}
\end{equation}
In the low energy limit $\xi(\sqrt{x}-1)\ll1$, one has
\begin{equation}
N(\gamma,t)\approx {q_0(p-2)t\over p}\gamma^{-p}\Biggl[1-\left({t_0\over t}\right)^{1/2}\Biggr]^2.
\label{eq:NSynSbL}
\end{equation}
One can obtain the high energy limit by setting $2\xi(\sqrt{x}-1)\to 1$ in (\ref{eq:NSynSb}), which leads to
$N(\gamma,t)\approx (q_0/2p)\tau_{s0}(\tau_{s0}/t_0)\gamma^{-p-2}$. The power index obtained here
is consistent with that derived in \citet{es89}. It is worth noting that their approximation gives 
an index $\sim p+1/(1-\alpha_B)$, which is not applicable for $\alpha_B=1$.

\subsection{General case}

Generally, the integral can be expressed in terms of the hypergeometric function
using the following integral
\begin{eqnarray}
\int^1_0(1+At)^\alpha(1+Bt)^\beta dt&=&
{(B-A)^\alpha\over (1+\beta)B^{1+\alpha}}\Biggl[
(1+B)^{1+\beta}F\Bigl(1+\beta,-\alpha;2+\beta;A{1+B\over A-B}\Bigr)
\nonumber\\
&&
-F\Bigl(1+\beta,-\alpha;2+\beta;{A\over A-B}\Bigr)\Biggr],
\end{eqnarray}
where the hypergeometric function is defined as \citep{as65}
\begin{equation}
F(a,b;c;z)={\Gamma(c)\over\Gamma(b)\Gamma(c-b)}\int^1_0t^{b-1}(1-t)^{c-b-1}(1-zt)^{-a}dt.
\label{eq:hyper2}
\end{equation}
One obtains
\begin{eqnarray}
I(\gamma_0,t)&=&
{[1-(1-\alpha_B)/\xi_0]^\alpha\over \xi_0(p-1)}
\Biggl[
(1+B)^{p-1}F\Bigl(p-1,-\alpha;p;{A(1+B)\over A-B}\Bigr)
\nonumber\\
&&-F\Bigl(p-1,-\alpha;p;{A\over A-B}\Bigr)\Biggr],
\label{eq:I}
\end{eqnarray}
where 
\begin{eqnarray}
A={x}^{1-\alpha_B}-1,\quad
B={\xi_0\over1-\alpha_B}\left({x}^{1-\alpha_B}-1\right),\quad
\alpha={\alpha_B\over1-\alpha_B}.
\end{eqnarray}
Again here the final solution can be obtained by substituting (\ref{eq:I}) into
(\ref{eq:NSynG}) and transforming it back to the $(\gamma,t)$ domain using (\ref{eq:transform}).
This gives
\begin{eqnarray}
N(\gamma,t)=q_0t_0\gamma^{-p}\Biggl[1-{\xi\over1-\alpha_B}\left(x^{1-\alpha_B}-1\right)\Biggr]^{p-2}I(\gamma,t), 
\label{eq:NSynG2}
\end{eqnarray}

Alternatively, the solution of the continuity equation (\ref{eq:CEq}) can be written as
\citep{k62}
\begin{eqnarray}
N(\gamma,t)=q_0\gamma^{-p}\!\int^t_{t_0}\Biggl[1-\gamma
\bar{\psi}(t,t') \Biggr]^{p-2}
\!\!\left({t\over t'}\right)^{(1-p)\alpha_V}\!\! dt',
\label{eq:NKard}
\end{eqnarray}
with
\begin{equation}
\bar{\psi}(t,t')\equiv\int^t_{t'}\left({t\over t''}\right)^{\alpha_V}{dt''\over\tau_1(t'')},
\label{eq:bpsi}
\end{equation}
where $\tau_1(t)$ is defined in (\ref{eq:b}). Eq (\ref{eq:bpsi})
can be integrated to yield
\begin{eqnarray}
\bar{\psi}=t\Biggl\{
{1\over(\alpha'-1)\tau_{s0}}\left({t\over t_0}\right)^{-\alpha_B}\!\!\Biggl[
\left({t\over t'}\right)^{\alpha'-1}-1\Biggr]
+{1\over(\alpha_V-1)\tau_{ICS}}\Biggl[\left({t\over
t'}\right)^{\alpha_V-1}-1\Biggr]\Biggr\}.
\label{eq:bpsi2}
\end{eqnarray}
The solution satisfies the initial condition $N(\gamma,0)=0$.
For $\alpha_B=1$, one has
\begin{equation}
\bar{\psi}(t,t')={t_0\over\tau_{s0}}\ln\left({t\over t'}\right).
\label{eq:bpsi3}
\end{equation}
Substituting  Eq (\ref{eq:bpsi3}) into (\ref{eq:NKard}) leads to (\ref{eq:NSyn}).

\end{appendix}

\newpage


\begin{thebibliography}{22}
\bibitem[\protect\citeauthoryear{Abramowitz \& Stegun}{1965}]{as65}
Abramowitz, M., Stegun, I. A., 1965, Handbook of Mathematical Functions (New York: Dover)
\bibitem[\protect\citeauthoryear{Becker, White \& Helfand}{1995}]{betal95}
Becker, R.H., White, R.L., Helfand, D.J. 1995, ApJ 450, 559 
\bibitem[\protect\citeauthoryear{Bicknell}{1986}]{b86}
Bicknell, G. V., 1986, ApJ, 300, 591
\bibitem[\protect\citeauthoryear{B\^irzan, et al.}{2008}]{betal08}
B\^irzan, L., McNamara, B. R., Nulsen, P. E., Carilli, C. L., Wise, M. W., 
2008, ApJ, 686, 859
\bibitem[\protect\citeauthoryear{Blundell, Rawlings \& Willott}{1999}]{betal99}
Blundell, K. M., Rawlings, S., Willott, C. J., 1999, ApJ, 117, 677
\bibitem[\protect\citeauthoryear{Bock, Large \& Sadler}{1999}]{bls99}
Bock, D.C.-J., Large, M.I., Sadler, E.M. 1999, AJ 117, 1578 
\bibitem[\protect\citeauthoryear{Cavaliere, Morrison \& Wood}{1971}]{cetal71}
Cavaliere, A., Morrison, P., Wood, K., 1971, ApJ, 170, 223
\bibitem[\protect\citeauthoryear{Canvin et al.}{2005}]{cetal05}
Canvin, J. R., Laing, R. A., Bridle, A. H., Cotton, W. D., 
2005, MNRAS, 363, 1223
\bibitem[\protect\citeauthoryear{Colless et al.}{2001}]{cetal01}
Colless, M. et al. 2001, MNRAS 328, 1039 
\bibitem[\protect\citeauthoryear{Condon et al.}{1998}]{cetal98}
Condon, J.J., Cotton, W.D., Greisen, E.W., Yin, Q.F., Perley, R.A., Taylor, G.B., Broderick, J.J.
1998,  AJ 115, 1693 
\bibitem[\protect\citeauthoryear{Condon}{1989}]{c89}
Condon, J. J., 1989, ApJ, 338, 13
\bibitem[\protect\citeauthoryear{Croston et al}{2008}]{cetal08}
Croston, J. H., Hardcastle, M. J., Birkinshaw, M., Worrall, D. M., Laing, R. A., 2008,
MNRAS, 386, 1709
\bibitem[\protect\citeauthoryear{De Young}{1980}]{d80}
De Young, D. S., 1980, ApJ, 241, 81
\bibitem[\protect\citeauthoryear{Eilek \& Shore}{1989}]{es89}
Eilek, J. A., Shore, S. N., 1989, 342, 187
\bibitem[\protect\citeauthoryear{Eilek, Melrose \& Walker}{1997}]{eetal97}
Eilek, J., Melrose, D. B., Walker, M. A., 1997, ApJ, 483, 282
\bibitem[\protect\citeauthoryear{Falle}{1991}]{f91}
Falle, S., 1991, MNRAS, 250, 581
\bibitem[\protect\citeauthoryear{Fanaroff \& Riley}{1974}]{fr74}
Fanaroff, B., Riley, J., 1974, MNRAS, 167, 31
\bibitem[\protect\citeauthoryear{Fedorenko \& Zentsova}{1986}]{fz86}
Fedorenko, V. N., Zentsova, A. S., 1986, Sov. Astron. 30, 24
\bibitem[\protect\citeauthoryear{Garrington \& Conway}{1991}]{gc91}
Garrington, S. T., Conway, R. G., 1991, MNRAS, 250, 198
\bibitem[\protect\citeauthoryear{Gopal-Krishna \& Wiita}{1987}]{gw87}
Gopal-Krishna, Wiita, P. J., 1987, MNRAS, 226, 531
\bibitem[\protect\citeauthoryear{Gopal-Krishna \& Wiita}{1988}]{gw88}
Gopal-Krishna, Wiita, P. J., 1988, Nature, 333,  49
\bibitem[\protect\citeauthoryear{Jackson \& Wall}{1999}]{jw99}
Jackson, C. A., Wall, J. V., 1999, MNRAS, 304, 160.
\bibitem[\protect\citeauthoryear{Jaffe \& Perola}{1973}]{jp73}
Jaffe, W. J., Perola, G. C., 1973, A\&A, 26, 423
\bibitem[\protect\citeauthoryear{Kaiser et al.}{1997}]{ketal97}
Kaiser, C. R. et al. 1997,  MNRAS, 292, 723
\bibitem[\protect\citeauthoryear{Kaiser \& Best}{2007}]{kb07}
Kaiser, C. R., Best, P. N., 2007, MNRAS, 381, 1548 
\bibitem[\protect\citeauthoryear{Kardashev}{1962}]{k62}
Kardashev, N. S., 1962, Soviet Astron--AJ, 6, 317.
\bibitem[\protect\citeauthoryear{Komissarov \& Falle}{1998}]{kf98}
Komissarov, S. S., Falle, S. A., 1998, MNRAS, 297, 1087.
\bibitem[\protect\citeauthoryear{Laing et al.}{2006}]{letal06}
Laing, R. A., Canvin, J. R., Bridle, A. H., Hardcastle, M., MNRAS, 372, 510
\bibitem[\protect\citeauthoryear{Ledlow \& Owen}{1996}]{lo96}
Ledlow, M. J., Owen, F. N., 1996, AJ, 112, 9
\bibitem[\protect\citeauthoryear{Longair}{1994}]{l94}
Longair, M., 1994, High Energy Astrophysics (Cambridge University Press)
\bibitem[\protect\citeauthoryear{Manolakou \& Kirk}{2002}]{mk02}
Manolakou, K., Kirk, J., 2002, A\&A, 391, 127
\bibitem[\protect\citeauthoryear{Mauch et al.}{2003}]{metal03}
Mauch, T., Murphy, T., Buttery, H.J., Curran, J., Hunstead, R.W., 
Piestrzynski, B., Robertson, J.G., Sadler, E.M. 2003, MNRAS
342, 1117 
\bibitem[\protect\citeauthoryear{Mauch \& Sadler}{2007}]{ms07}
Mauch, T., Sadler, E. M., 2007, MNRAS, 375, 931
\bibitem[\protect\citeauthoryear{Melrose}{1980}]{m80}
Melrose, D. B., 1980, Plasma Astrophysics, Vol. 2
\bibitem[\protect\citeauthoryear{Mulchaey \& Zabludoff}{1998}]{mz98}
Mulchaey, J. S., Zabludoff, A. I., 1998, ApJ, 496, 73
\bibitem[\protect\citeauthoryear{Myers \& Spangler}{1985}]{ms85}
Myers, S. T., Spangler, S. R., 1985, 291, 52
\bibitem[\protect\citeauthoryear{Parma et al.}{2002}]{petal02}
Parma, P., Murgia, M., de Ruiter, H. R., Fanti, R., 2002, New Astron. Rev. 46, 313
\bibitem[\protect\citeauthoryear{Sadler et al.}{2002}]{setal02}
Sadler, E. M. et al. 2002, MNRAS, 329, 227
\bibitem[\protect\citeauthoryear{Scheuer}{1974}]{s74}
Scheuer, P. A. G., 1974, MNRAS, 166, 513
\bibitem[\protect\citeauthoryear{Schmidt}{1966}]{s66}
Schmidt, M., 1966, ApJ, 146, 7
\bibitem[\protect\citeauthoryear{Shklovskii}{1963}]{s63}
Shklovskii, I. S., 1963, Soviet Astron--AJ, 6, 465
\bibitem[\protect\citeauthoryear{Slee et al.}{1994}]{setal94}
Slee, O.B., Sadler, E.M., Reynolds, J.E., Ekers, R.D. 1994, MNRAS 269, 928 
\end{thebibliography}
\end{document}